\documentclass[preprint,showpacs,preprintnumbers,amsmath,amssymb]{revtex4-1}
\usepackage{graphicx}
\usepackage{amsmath}
 \usepackage{epstopdf}
\usepackage{color}

\usepackage{color}

\begin{document}
\title{Long-term and seasonal variability of wind and wave extremes in the Arctic Ocean}

\author{I. S. Cabral$^{1}$}
\author{I. R. Young$^1$}
\author{A. Toffoli$^{1}$}

\affiliation{
$^1$Department of Infrastructure Engineering, The University of Melbourne, Parkville, VIC 3010, Australia
}

\date{\today}

%

\begin{abstract}
Over recent decades, the Arctic Ocean has experienced dramatic changes due to climate change. Retreating sea ice has opened up large areas of ocean, resulting in an enhanced wave climate. Taking into account the intense seasonality and the rapid changes to the Arctic climate, a non-stationary approach is applied to time-varying statistical properties to investigate historical trends of extreme values. The analysis is based on a 28-year wave hindcast (from 1991 to 2018) carried out with the WAVEWATCH III wave model forced by ERA5 wind speed. The results show notable seasonal differences and robust positive trends in extreme wave height and wind speed, especially in the Beaufort and East Siberian seas, with increasing rates in areal-average of the 100-year return period of wind speed of approximately 4\% and significant wave height up to 60\%. It is concluded that the significant increases in extreme significant wave height are largely associated with sea-ice retreat and the enhanced fetches available for wave generation.
\end{abstract}

\maketitle

\section{Introduction}

\begin{figure}
  \centering
  \includegraphics[scale=0.4]{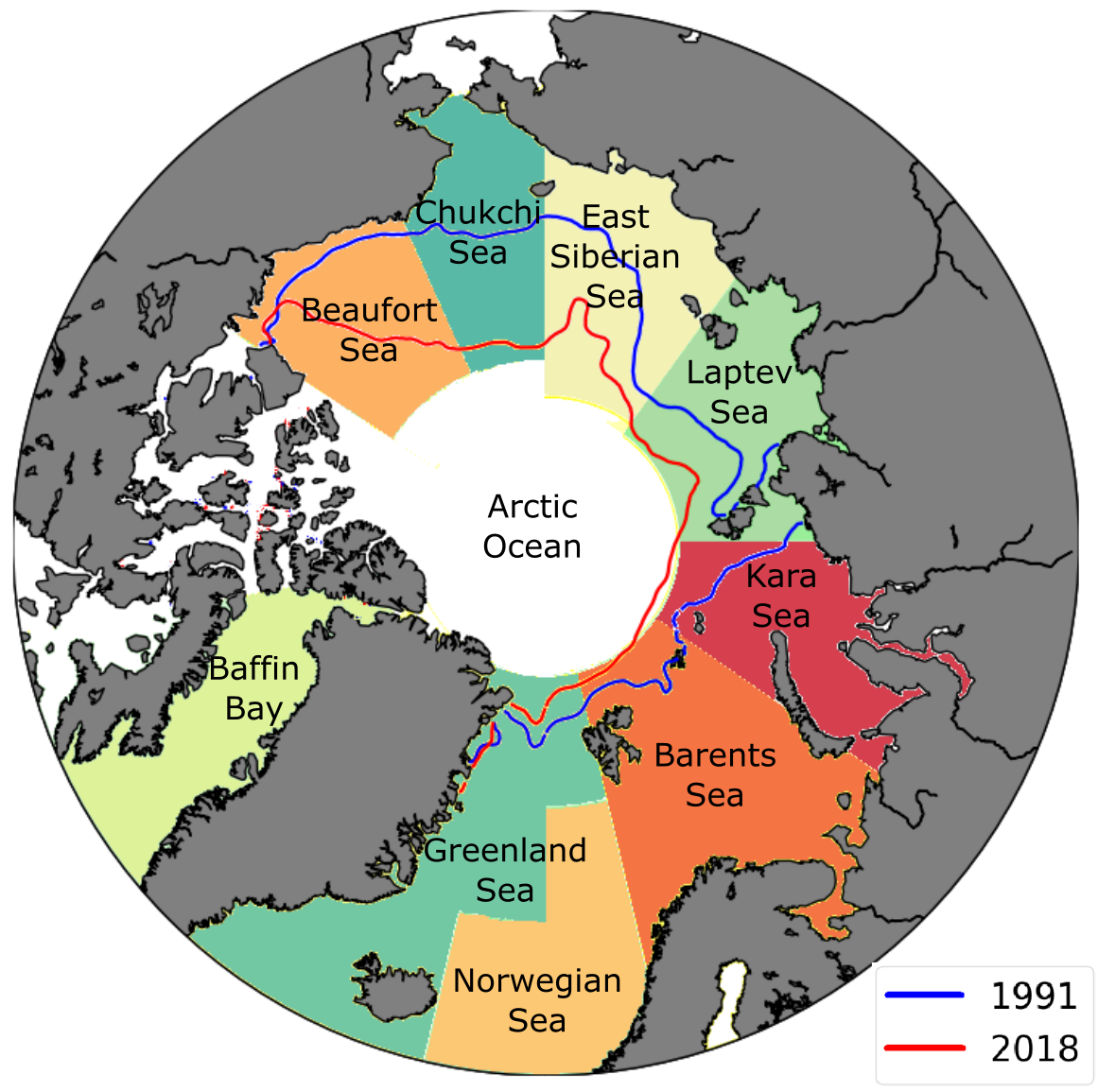}
  \caption{Regions of the Arctic Ocean used in this study with lines showing sea ice extent in September of 1991 (blue) and 2018 (red). Sea ice concentration dataset from ERA5 reanalysis.}
  \label{arctic-seas}
\end{figure}

Arctic sea ice extent has been declining sharply for the past three decades (see minimum sea ice extent in September 1991 and September 2018 in Fig. \ref{arctic-seas}). Variations of the sea ice cover has been the cause of notable changes to meteorological and oceanographic conditions in the Arctic Ocean \citep[e.g.][]{Thomson2014,Stopa2016a,Liu2016,Thomson2016,Waseda2018,casassea}. Emerging open waters provide longer fetches for surface waves to build up more energy and increase in magnitude \citep{Thomson2014,Thomson2016}. Concurrently, an increase of wave height impacts profoundly on the already weak sea ice cover by enhancing breakup and melting processes in a feedback mechanism \citep{Thomson2016,dolatshah2018hydroelastic,passerotti2020omae}. In addition, coastlines and coastal communities have been impacted by intensifying erosion with coastline retreat rates up to 25\,m per year \citep[e.g.][]{Jones2009,gunther2015observing}.

Ocean climate evaluated from satellite observations \citep{Liu2016} for the months of August and September---the period of minimum ice coverage---reveal weak or even negative trends of average offshore wind speeds over the period between 1996 and 2015, while notable upward trends were detected in the higher $90^{th}$ and $99^{th}$ percentiles across the entire Arctic Ocean, except for the Greenland Sea. Unlike winds, waves showed more substantial increasing rates even for average values, especially in the Chukchi, Laptev, Kara seas and Baffin Bay. 

Satellite observations have temporal and spatial limitations, which are exacerbated in the Arctic where most of the altimeter sensors do not usually cover latitudes higher than $82^{\circ}$. Numerical models, on the contrary, provide more consistent data sets for climate analysis in this region. \citet{Stopa2016a} estimated trends using a 23-year model hindcast and found that simulated average wind speed exhibits a weak increasing trend, especially in the Pacific sector of the Arctic Ocean, slightly differing from the satellite-based observations in \citet{Liu2016}. Average wave heights, however, were found to be consistent with altimeter observations. \cite{Waseda2018} used the ERA-Interim reanalysis database \citep{Dee2011} to evaluate the area-maximum wind speed and wave height in the months of August, September and October from the period 1979-2016 in the Laptev, East Siberian, Chukchi, and Beaufort seas. Their analysis indicated robust increasing trends for both variables, with most significant changes in October: $\approx 0.06 \;$ m/s per year for wind speed and $\approx 2\;$cm per year for mean significant wave height. Recently, \citet{casassea} simulated historical (1979-2005) and future (2081-2100) sea state conditions to evaluate changes in regional annual maximum significant wave height, under high baseline emission scenarios (RCP8.5). Their results indicated that wave height is projected to increase at a rate of approximately 3 cm per year, which is more than 0.5\% per year.

Previous assessment of ocean climate in the Arctic has focused on annual or monthly values and often paid specific attention to summer months. A comprehensive evaluation of climate and related changes, however, cannot ignore properties and frequency of occurrence of extremes. Classically, the latter is estimated with an extreme value analysis (EVA), where observations are fitted to a theoretical probability distribution to extrapolate values at low probability levels, such as those occurring on average once every 100 years \citep[normally referred to as the 100-year return period event,][]{thomson2014data}. Therefore, the EVA has to rely on long records spanning over one or more decades (observations typically cover more than a 1/3 of the return period), to be statistically significant. Furthermore, the EVA relies on the fundamental assumption that the statistical properties of the variable do not change over time, namely the process is stationary. For the strongly seasonal and rapidly changing Arctic environment, however, the hypothesis of stationarity cannot hold for an extended period of time, invalidating the fundamental assumption of the EVA. 

An alternative approach that better fits the highly dynamic nature of the Arctic is the estimation of time-varying extreme values with a non-stationary analysis \citep[see, for example,][for a general overview]{Coles2001,Mendez2006,galiatsatou2011modeling,cheng2014non,Mentaschi2016}. There are few methods for the estimation of time-varying extreme value distributions from non-stationary time series. A functional approach is the transformed-stationary extreme value analysis (TS-EVA) proposed by \citet{Mentaschi2016}. The method consists of transforming a non-stationary time series with a normalisation based on the time-varying mean and standard deviation into a stationary counterpart, for which the classical EVA theory can be applied. Subsequently, an inverse transformation allows the conversion of the EVA results in time-varying extreme values. 

Here we apply the TS-EVA method to assess time-varying extremes in the Arctic Ocean. The assessment is performed on a data set consisting of a long-term hindcast---from January 1991 to December 2018---that was obtained using the WAVEWATCH III \citep[WW3,][]{Tolman2009} spectral wave model forced with ERA5 reanalysis wind speeds \citep{hersbach2019global}.  A description of the model and its validation is reported in Section \ref{ww3}. Model data are processed with the TS-EVA to determine extreme values for wind forcing and wave height. Long-term trends are investigated with a nonseasonal approach, and seasonal variability considered with a concurrent seasonal approach (Section \ref{TSEVA}). Results are discussed in terms of regional distributions and areal averages in Sections \ref{nses} and \ref{ses}. Concluding remarks are presented in the last Section.

\section{Wave hindcast}\label{ww3}


A 28-year (from 1991 to 2018) wave hindcast of the Arctic Ocean was carried out with the WAVEWATCH III (WW3) spectral wave model---version 6.07---to build a database of sea state conditions, which is consistent in space and time. A regional model domain covering the area above latitude $53.17^{\circ}$ N was set up in an Arctic Polar Stereographic Projection with a horizontal resolution varying from 9 to 22 km (this configuration was found to optimise the accuracy of model results in relation to recorded data and computational time). The bathymetry was extracted from the ETOPO1 database \citep{amante2009etopo1}. The regional set up was then forced with ERA5 atmospheric forcing and sea ice coverage \citep{hersbach2019global}. Note that the model ran without wave-ice interaction modules as the focus is on the open ocean and not the marginal ice zone; regions of sea ice with concentration larger than 25\% were therefore treated as land \citep[e.g.][]{Thomson2016}. The model physics were defined by the ST6 source term package \citep{Zieger2015}. Boundary conditions were imposed on the regional model to account for energetic swells coming from the North Atlantic. To this end, boundaries were forced by incoming sea states from WW3 global runs with 1-degree spatial resolution. The global model used ERA5 wind forcing and the ST6 source term package. Simulations were run with a spectral domain of 32 frequency and 24 directional bins (directional resolution of 15 degrees). The minimum frequency was set at 0.0373 Hz and the frequency increment factor was set at 1.1, providing a frequency range of 0.0373-0.715 Hz. Grid outputs were stored every 3 hours. 

Calibration of the wind-wave growth parameter (CDFAC) was performed by testing the model outputs (significant wave height) against altimeter data across six different satellite missions \citep[ERS1, ERS2, ENVISAT, GFO, CRYOSAT-2 and Altika SARAL, see][]{Queffeulou2015} for the period August-September 2014. Note that the calibration of the regional configurations was undertaken after tuning the global model, to allow the input of reliable boundary conditions in the former. The best agreement was achieved for CDFAC = 1.19 in the global model, with correlation coefficient $R=0.96$, scatter index $SI=16\%$ and root mean square error $RMSE = 0.4$ m. For the regional model, the best agreement was for CDFAC = 1.23 with $R=0.95$, $SI\sim1\%$ and $RMSE \sim 0.3$ m. 

\begin{figure}
  \centering
  \includegraphics[scale=0.6]{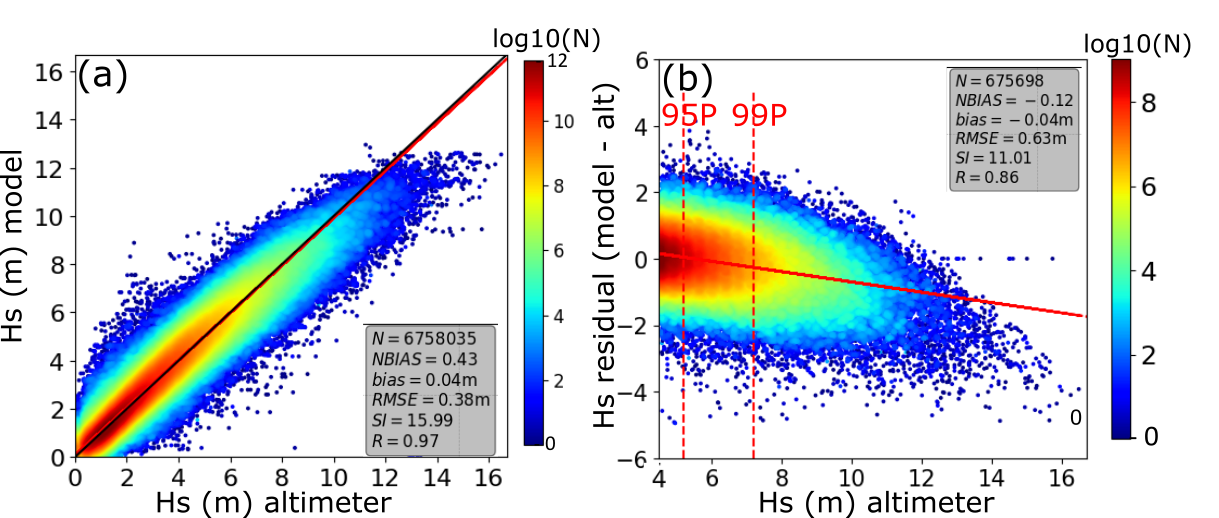}
  \caption{Significant wave height from model versus collocated altimeter observations for the period 2012--2016 with ST6 core physics. (a) all data and (b) $90^{th}$ percentile and above. The black line represents the 1:1 agreement and the red lines are the linear regression.}
  \label{validationst4st62}
\end{figure}

\begin{figure}
  \centering
  \includegraphics[scale=0.45]{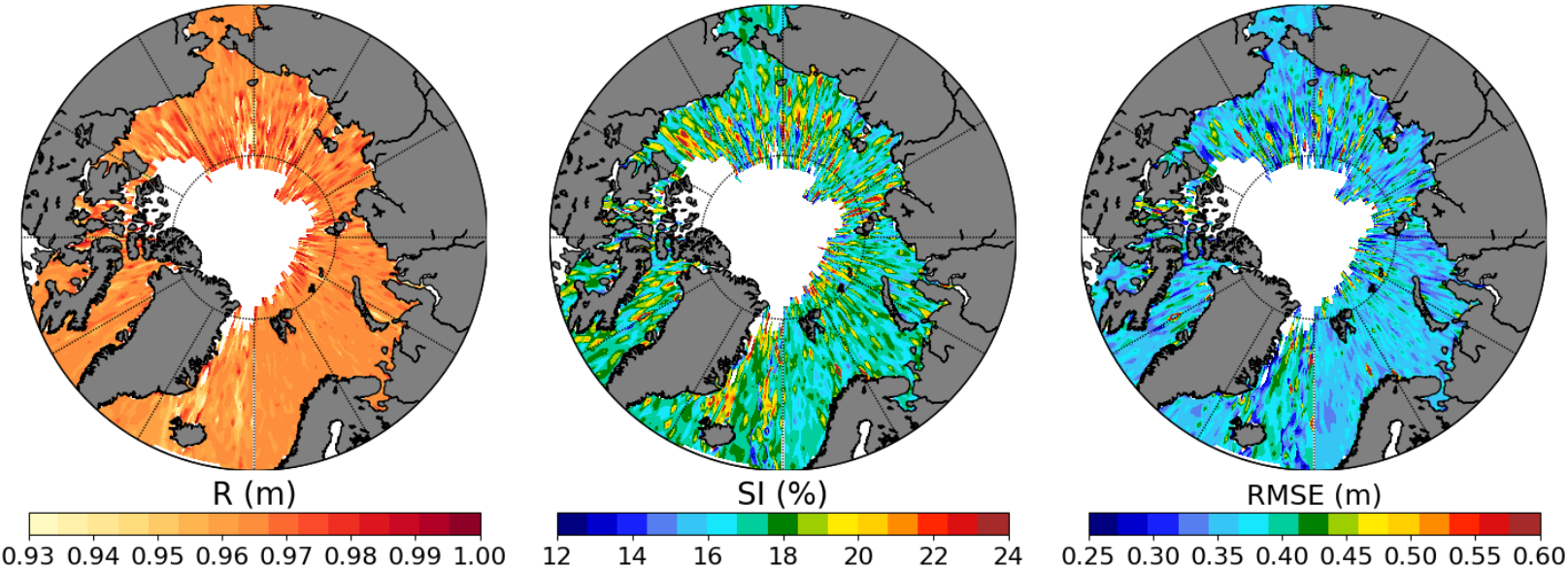}
  \caption{Regional distribution of error metrics: correlation (left panel), scatter index (middle panel), and root mean square error (right panel).}
  \label{validation-map}
\end{figure}

The regional model set up was further validated by comparing all modelled significant wave height values against matching altimeter observations for an independent period of four years from 2012 to 2016. Fig. \ref{validationst4st62}a shows the regional model outputs versus collocated altimeter data. Generally, the model correlates well with observations: $R = 0.97$, $SI = 16\%$, and $RMSE = 0.38$ m. The residuals between model and altimeters as a function of the observations are reported in Fig. \ref{validationst4st62}b. The comparison indicates a satisfactory level of agreement for the upper range of wave heights ($H_s>4$ m): $R = 0.86$, $SI = 11\%$, and $RMSE = 0.63$ m. The regional distribution of model errors is reported in Fig. \ref{validation-map}. It is worth noting that the model performed well across the entire Arctic Ocean, with no specific regions affected by significant errors. Further evaluation of the performance of the wave hindcast against ERA5 wave reanalysis is described in Appendix A.

\section{Transformed Stationary Extreme Value Analysis (TS-EVA)} \label{TSEVA}

\begin{figure}
  \centering
  \includegraphics[scale=0.1]{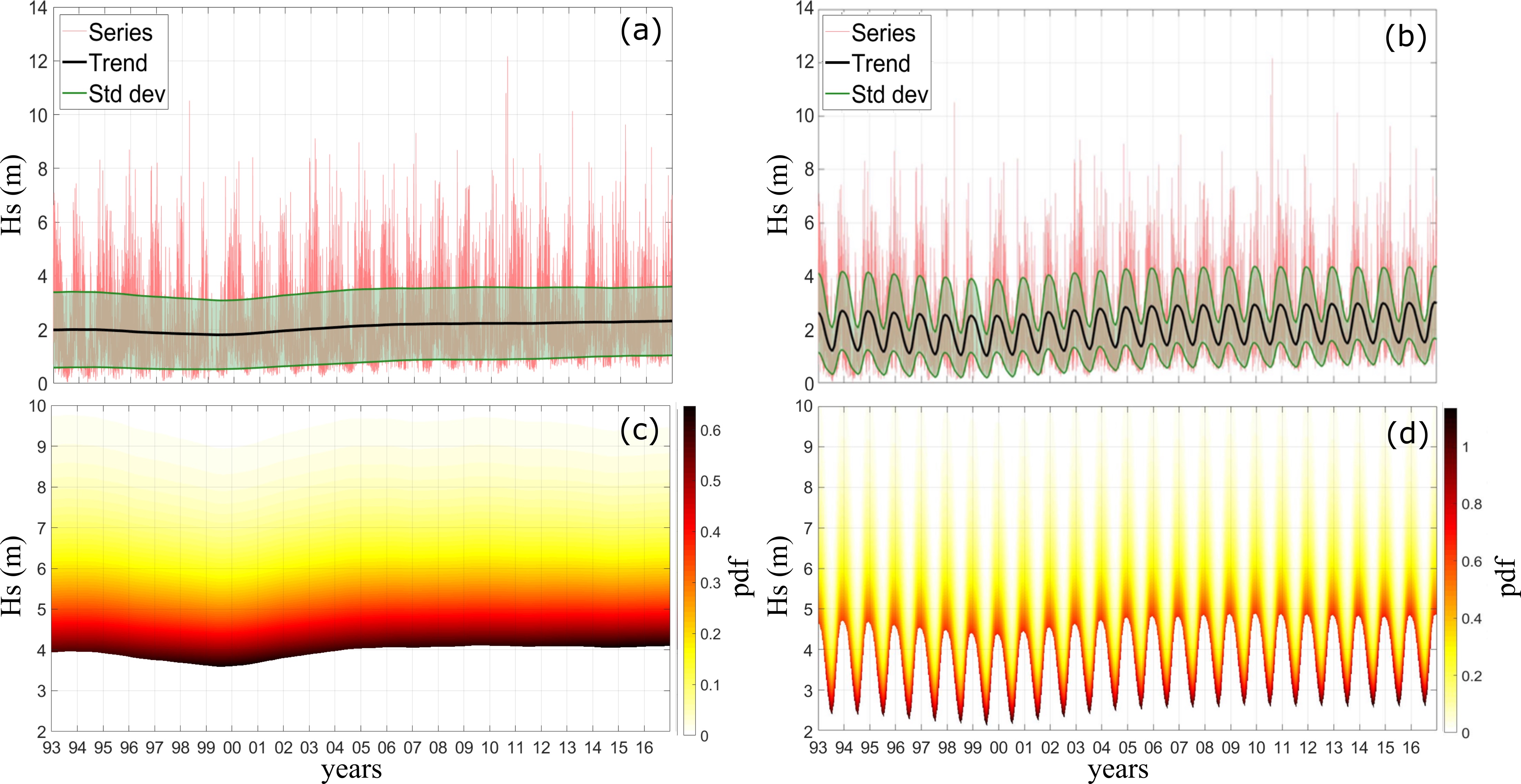}
  \caption{TS-EVA of the projections of significant wave height for a point located in the Kara Sea. The time series of $H_s$ (m), its long-term trend and standard deviation computed with a time window of 5 years obtained with (a) the nonseasonal approach and (b) with the seasonal approach. The non-stationary time-dependent probability distribution for a GPD with a POT analysis and a $90^{th}$ percentile threshold with (c) the nonseasonal approach and (d) with the seasonal approach.}
  \label{ts-kara-nons}
\end{figure}

The TS-EVA method developed by \cite{Mentaschi2016} is applied to extract time-varying information on climate extremes. This approach is based on three main steps. In the first step, the original non-stationary time series is transformed into a stationary counterpart that can be processed using classical EVA methods. The transformation is based on the following equation:
\begin{equation} 
x(t)=\frac{y(t)-T_y (t)}{S_y (t)},
\label{eq63}
\end{equation} 
where $y(t)$ is the non-stationary time-series, $x(t)$ is the stationary counterpart, $T_y (t)$ is the trend of $y(t)$ and the $S_y (t)$ is its standard deviation. Computation of $T_y (t)$ and $S_y (t)$ relies on algorithms based on running means and running statistics \citep[see][for more details]{Mentaschi2016}. This approach acts as a low-pass filter, which removes the variability within a specified time window $W$. The time window has to be short enough to incorporate the desired variability, but long enough to eliminate noise and short-term variability. Hereafter this approach is referred to as nonseasonal. A period of 5 years for $W$ is used to ensure stationary transformed time series, considering the rapid sea ice melting occurring in the last few decades in the region. Fig. \ref{ts-kara-nons}a shows an example of a time-series of significant wave height for the Kara sea, its long-term variability and concurrent standard deviation. Apart from an initial downward trend between 1993 and 1999, when the region was still covered by sea ice for most of the year, a clear positive trend is evident for the past two decades. 

In the second step, the stationary time-series $x(t)$ is processed with a standard EVA approach. Herein, a peaks-over-threshold method (POT) \citep[see, e.g.][for a general overview]{thomson2014data} was applied to extract extreme values from the records with a threshold set at the $90^{th}$ percentile. A Generalised Pareto Distribution \citep[GPD, e.g.][]{thomson2014data}

\begin{equation} 
F(x)=1-\Bigg[1+k\Bigg(\frac{x-A}{B}\Bigg)^\frac{-1}{k}\Bigg],
\label{eqgpd}
\end{equation}
where $A$ is the threshold and $B$ and $k$ are the scale and shape parameters respectively, was fitted to the data in order to derive an extreme value distribution. Note that the parameters $A$ and $B$ are time-dependent and change with trends, standard deviation, and seasonality in the TS-EVA approach \citep{Mentaschi2016}. To ensure statistical independence, peaks were selected at least 48 hours apart. Furthermore, to ensure a stable probability distribution, a minimum of 1000 peaks was selected for each grid point of the model domain \citep{Meucci2018WindEnsembles}, meaning that regions free of sea ice less than about two months per year were excluded from the analysis. 

The third and final step consists of back-transforming the extreme value distribution into a time-dependent one by reincorporating the trends that were excluded from the original non-stationary time series. An example of the non-stationary extreme value distribution for a point located in the Kara Sea is shown in Fig. \ref{ts-kara-nons}c. As the resulting distribution is different for each year within the time series, the TS-EVA method enables extrapolation of partial return period values (e.g. the 100-year return level for wind speed and significant wave height) for any specific year. Therefore, after fitting a GPD distribution to the stationary time series and transforming to a time-varying distribution, it was possible to obtain 100-year return levels for every five years within the original time series.

Effects of the seasonal cycle can be accounted for by incorporating seasonal components in the stationary time-series $x(t)$. To this end, trend $T_y (t)$ and standard deviation $S_y (t)$ in equation (\ref{eq63}) are expressed as $T_y(t)=T_{0y}(t)+s_T(t)$ and $S_y(t)=S_{0y}(t)\times s_S(t)$, where $T_{0y}(t)$ and $s_T(t)$ are the long-term and seasonal components of the trend and $S_{0y}(t)$ and $s_S(t)$ are the long-term and seasonal components of the standard deviation. $T_{0y}(t)$ and $S_{0y}(t)$ are computed by a running mean acting as a low-pass filter within a given time window ($W$). The seasonal component of the trend $s_T(t)$ is computed by estimating the average monthly anomaly of the de-trended series. The seasonal component of the standard deviation $s_S(t)$ is evaluated as the monthly average of the ratio between the fast and slow varying standard deviations, $S_{sn}(t)$ / $S_{0y}(t)$, where $S_{sn}$ is computed by another running mean standard deviation on a time window $W_{sn}$ much shorter than one year  \citep[see][for more details]{Mentaschi2016}. As for the non-seasonal approach, the time window $W$ was set to 5 years to estimate the long-term components, while a time window $W_{sn}$ of 2 months was applied to evaluate the intra-annual variability (seasonal components). Note that the length of the seasonal window $W_{sn}$ is chosen to maximise accuracy and minimise noise. As an example, Fig. \ref{ts-kara-nons}b shows the seasonal components for the Kara sea. The resulting stationary time series $x(t)$ is analysed with an EVA approach to fit an extreme value distribution, which is then back-transformed to a time-dependent one (Fig. \ref{ts-kara-nons}d). The seasonal approach enables the extrapolation of partial extreme values such as the 100-year return period levels for each month.

\section{Nonseasonal trends}\label{nses}

\subsection{Wind extremes}

\begin{figure}
  \centering
  \includegraphics[scale=0.7]{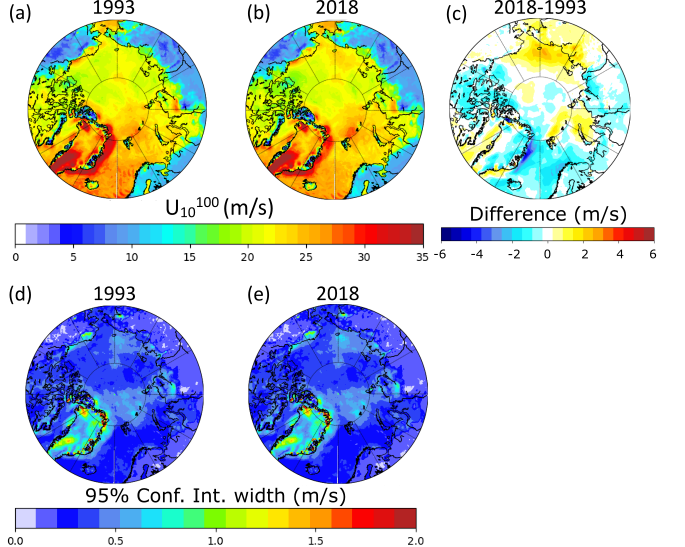}
  \caption{$U_{10}^{100}$  (m/s) obtained with a POT analysis ($90^{th}$ percentile threshold) and a GPD distribution in the TS-EVA nonseasonal approach for (a) 1993 and (b) 2018. (c) The difference between estimations for 2018 and 1993. Width of 95\% confidence interval for $U_{10}^{100}$ for (d) 1993 and (e) 2018.}
  \label{ci-nonseasonal-u10}
\end{figure}

\begin{figure}
  \centering
  \includegraphics[scale=0.17]{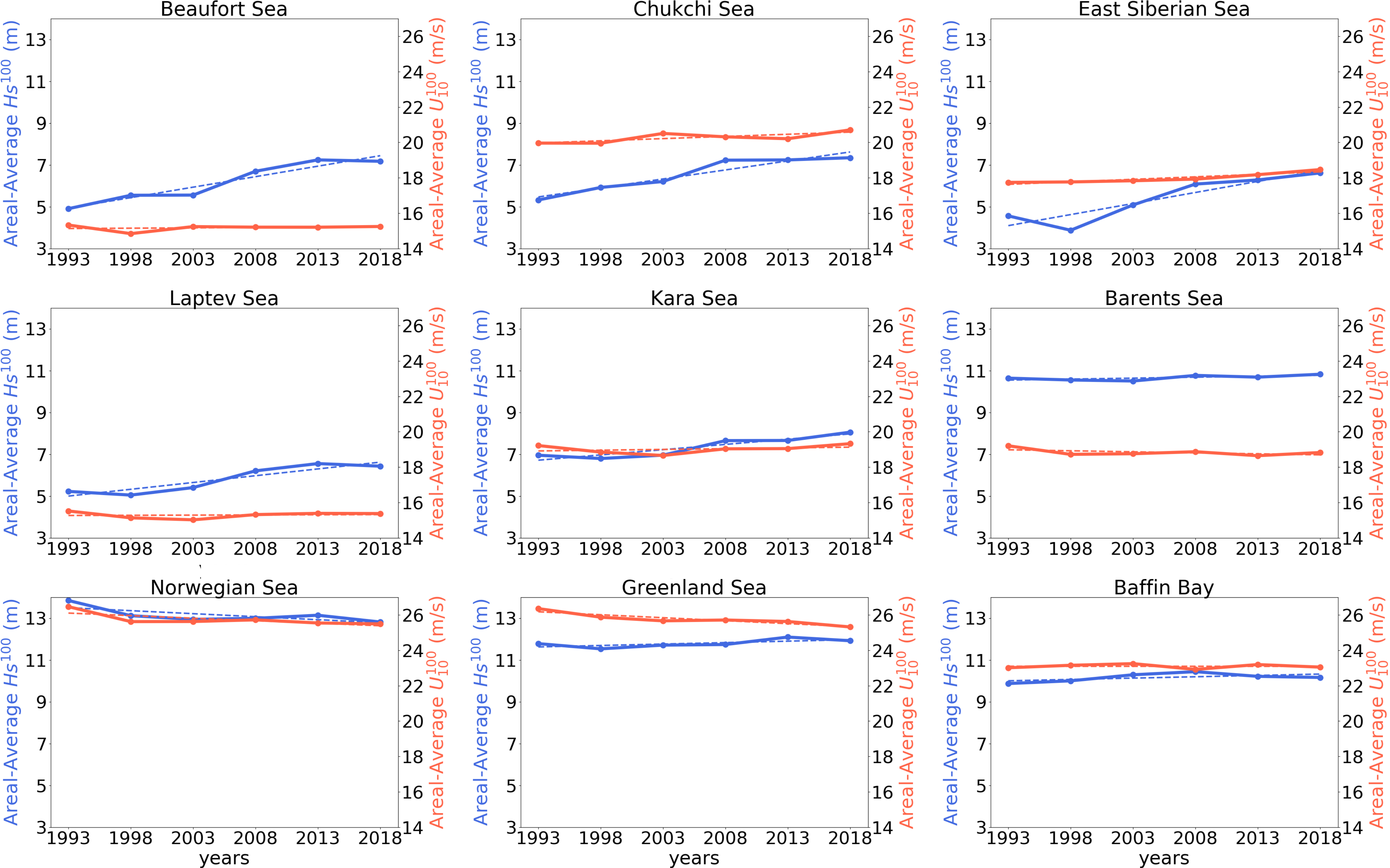}
  \caption{Areal-averages of $H_s^{100}$ (blue) in meters and $U_{10}^{100}$ (red) in $m/s$ estimated by nonseasonal TS-EVA approach for each sea in the Arctic Ocean.}
  \label{area-hs}
\end{figure}

Fig. \ref{ci-nonseasonal-u10} shows the regional distribution of the 100-year return period levels for wind speed $U_{10}^{100}$ and 95\% confidence interval (CI95) width for the years 1993 and 2018; the regional distribution of the differences between the two years is also displayed in the figure. Extreme winds are estimated to reach approximately 25 m/s in the Baffin Bay, Greenland, Barents and Kara seas (i.e. the Atlantic sector of the Arctic Ocean, see Fig. \ref{arctic-seas} for the geographical location of sub-regions), with peaks up to 40\;$m/s$ along the Eastern coast of Greenland. Extreme winds in the Pacific Sector, i.e. the Beaufort, Chukchi, East Siberian and Laptev seas recorded slightly lower $U_{10}^{100}$, reaching values up to 20 m/s. Confidence intervals were normally narrow, with extremes varying within the range of $\pm0.5$ m/s. The magnitude of extreme wind speeds predicted here is generally consistent with values determined with classical EVA methods in the Atlantic sector of the Arctic Ocean \citep{Breivik2014WindEnsembles,bitner2018climate}. The TS-EVA analysis, nevertheless, shows that extremes have been changing for the past three decades. The difference in the 100-year return period wind speeds between years 1993 and 2018 are notable as shown in Fig. \ref{ci-nonseasonal-u10}. More specifically, the long term trends of $U_{10}^{100}$ are shown in Fig. \ref{area-hs}, which reports areal-averages as a function of time for each sub-region. In the Atlantic sector, $U_{10}^{100}$ showed a weak drop in the Norwegian and Greenland seas, with a total decrease of about 3 m/s over the period 1993-2018 (a rate of -0.12 m/s per year). More significant drops were recorded along the Western coast of Greenland (i.e. Fram Strait, Eastern Greenland sea), where $U_{10}^{100}$ reduced at a rate of -0.24 m/s per year. The Baffin Bay and the Barents sea showed negligible changes, with $U_{10}^{100}$ remaining approximately constant. The opposite trend was reported on the Eastern side of the Atlantic sector (i.e. the Kara sea), where wind speed showed a weak increase with a rate of 0.04 m/s per year. The Pacific sector, on the contrary, was subjected to more consistent trends across the sub-regions. The East Siberian and Chukchi seas show weak positive trends of about 0.16 and 0.12 m/s per year, respectively. A similar increase was also observed in the Western part of the Beaufort sea. The Laptev sea recorded the lowest rate of increase in the Pacific sector, with $U_{10}^{100}$ increasing at a rate of 0.04 m/s per year.

\subsection{Wave extremes}

\begin{figure}
  \centering
  \includegraphics[scale=0.7]{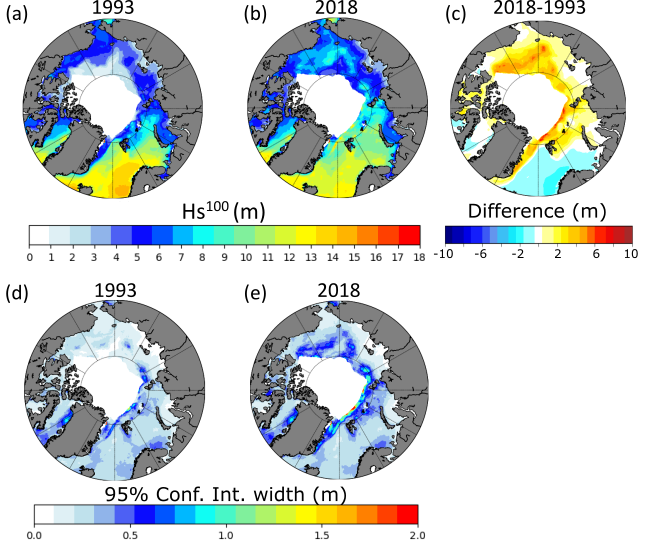}
  \caption{$H_s^{100}$ (m) obtained with a POT analysis ($90^{th}$ percentile threshold) and a GPD distribution in the TS-EVA nonseasonal approach for (a) 1993 and (b) 2018. (c) The difference between estimations for 2018 and 1993. Width of 95\% confidence interval for $H_s^{100}$ for (d) 1993 and (e) 2018.}
  \label{ci-nonseasonal}
\end{figure}

Fig. \ref{ci-nonseasonal} shows the 100-year return levels for significant wave height ($H_s^{100}$), confidence intervals and differences for the years 1993 and 2018. It should be noted that regions covered by sea ice for most of the year are not considered in this analysis and thus they are colour-coded with white in the figure. The Atlantic sector experiences high $H_s^{100}$ ($>10$ m) due to the energetic North Atlantic swell penetrating the Arctic Ocean. Likewise, the Pacific sector experiences significant values of $H_s^{100}$ ($>5$ m), despite a substantial sea ice cycle that limits fetch lengths for a large fraction of the year. The 95\% confidence intervals are typically $\pm0.5$ m (see panels d and e in Fig. \ref{ci-nonseasonal}). In more recent years (e.g. 2018), confidence intervals widen slightly in regions of significant sea ice decline, increasing to $\pm 0.6$ m.    

There is a clear difference of $H_s^{100}$ between 1993 and 2018, which appears consistent with the measured sea ice decline. There is a substantial increase of $H_s^{100}$ in the Pacific sector, with $H_s^{100}$ increasing by approximately 4 m in the Beaufort, Chukchi and East Siberian seas. In the Laptev and Kara seas, differences are typically smaller (the increment is approximately 2~m), even though $H_s^{100}$ reaches values of approximately 6 m nearby the sea ice margins. Note, however, that uncertainties related to the exact position of sea ice edges result in larger confidence intervals (up to $\pm2$~m) in these regions. Extremes in the Atlantic sector, surprisingly, show an overall decrease, with $H_s^{100}$ dropping by about 1-2~m. Note, however, that this is a region in which the sea ice extent has not changed dramatically over this period. Nevertheless, regions closer to sea ice such as the Fram straits and the Northern part of the Barents sea experienced substantial growth, with $H_s^{100}$ increasing up to 5~m between 1993 and 2018.

Trends in $H_s^{100}$ are reported in Fig. \ref{area-hs}. A consistent increase of $H_s^{100}$ is evident in the emerging open waters of the Beaufort, Chukchi, East Siberian, Laptev and Kara seas. Variations in the Beaufort and East Siberian seas are the largest, with a total increase over the period 1993-2018 of approximately 16~cm per year. The Chukchi and Laptev seas also experienced a substantial growth of $H_s^{100}$, with an increase of 6~cm per year, while $H_s^{100}$ increased by approximately 4~cm per year in the Kara sea. In contrast, the Atlantic sector shows only weak upward trends, with the Baffin Bay and Greenland sea showing an increase of 1.6~cm per year. The Barents sea experienced no notable long-term variations, while the Norwegian sea reported a drop in $H_s^{100}$ of about 4~cm per year. As these latter regions are predominantly free from sea ice, the downward trends are associated with the decline of wind speeds over the North Atlantic \citep[results are consistent with finding in][]{Breivik2013WaveForecasts,bitner2018climate}. It is worth noting that negative trends for the North Atlantic are expected to continues in the future as indicated by projections based on RCP 4.5 and RCP 8.5 emission scenarios \citep{morim2019robustness,aarnes2017projected}. Wave height, however, is predicted to increase at high latitudes of the Norwegian and Barents seas over the next decades as a result of ice decline \citep{aarnes2017projected}, confirming the positive trend in wave extremes that is already arising close the ice edge (see Fig. \ref{ci-nonseasonal}). The contrast between an overall decrease of wave height as a result of wind speed decline and the increase of wave height in emerging open waters at high latitudes is also a distinct feature in the North Pacific \citep[cf.][]{shimura2016variability}. 

\begin{figure}
  \centering
  \includegraphics[scale=0.5]{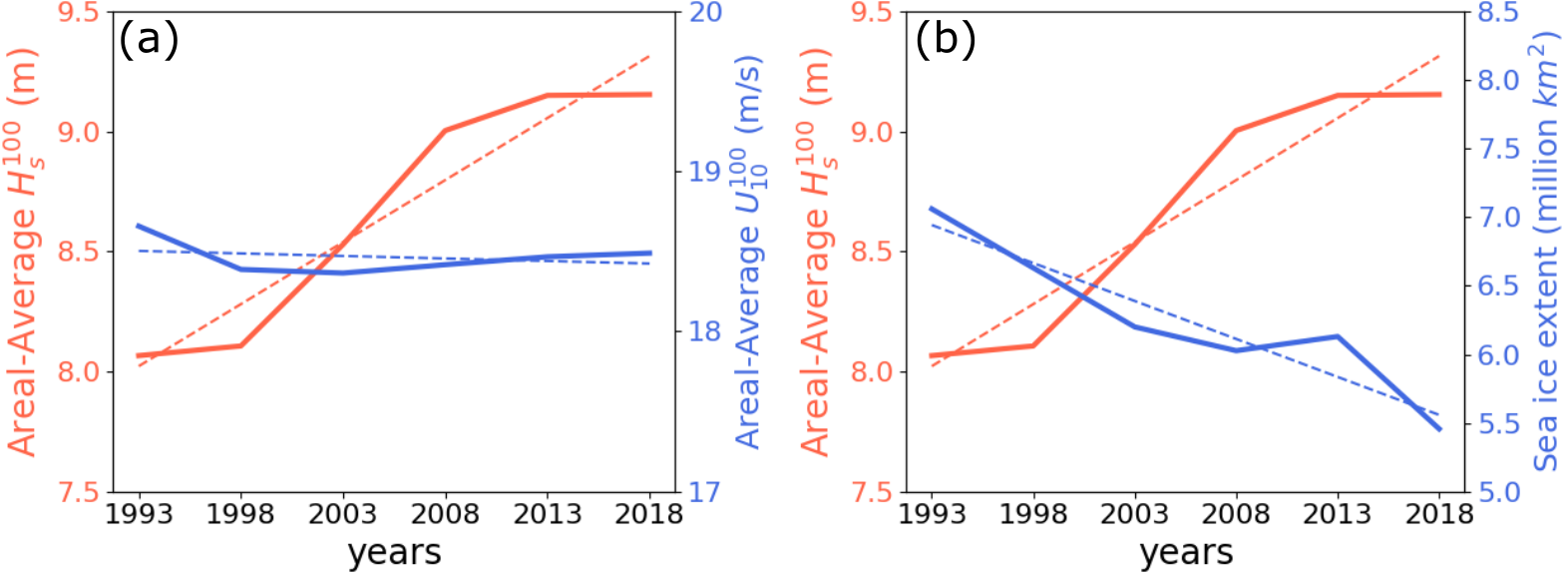}
  \caption{(a) Areal-average of $H_s^{100}$ in meters across the entire Arctic Ocean against areal-average of $U_{10}^{100}$ in m/s and (b) against sea ice extent in million km$^2$. Areal trend shown with the dashed lines.}
  \label{comparehsu10ice}
\end{figure}

Areal averages for $H_s^{100}$, $U_{10}^{100}$ and sea ice extent across the entire Arctic Ocean are shown in Fig. \ref{comparehsu10ice} as a function of time. Fig. \ref{comparehsu10ice}a confirms that the weak trends in wind extremes do not fully substantiate the significant changes in wave extremes. However, the substantial contraction of sea ice cover (about $27\%$ in 25 years) exhibits a more robust correlation with trends of wave extremes, corroborating that the emergence of longer fetches, i.e. sea ice decline, contributes notably to the positive trends of $H_s^{100}$ in the Arctic Ocean (Fig. \ref{comparehsu10ice}b).

\section{Seasonal variability}\label{ses}

\subsection{Wind extremes}

\begin{figure}
  \centering
  \includegraphics[scale=0.7]{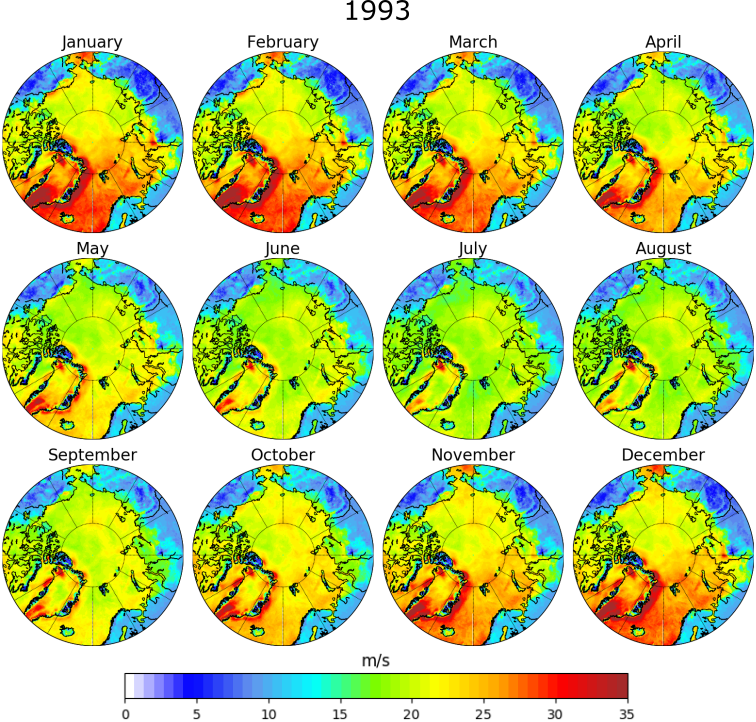}
  \caption{$U_{10}^{100}$ (m/s) for 1993 obtained with a POT analysis and a GPD distribution in the TS-EVA seasonal approach. Data obtained from the ERA5 dataset.}
  \label{u10-seasonal-1993}
\end{figure}

\begin{figure}
  \centering
  \includegraphics[scale=0.7]{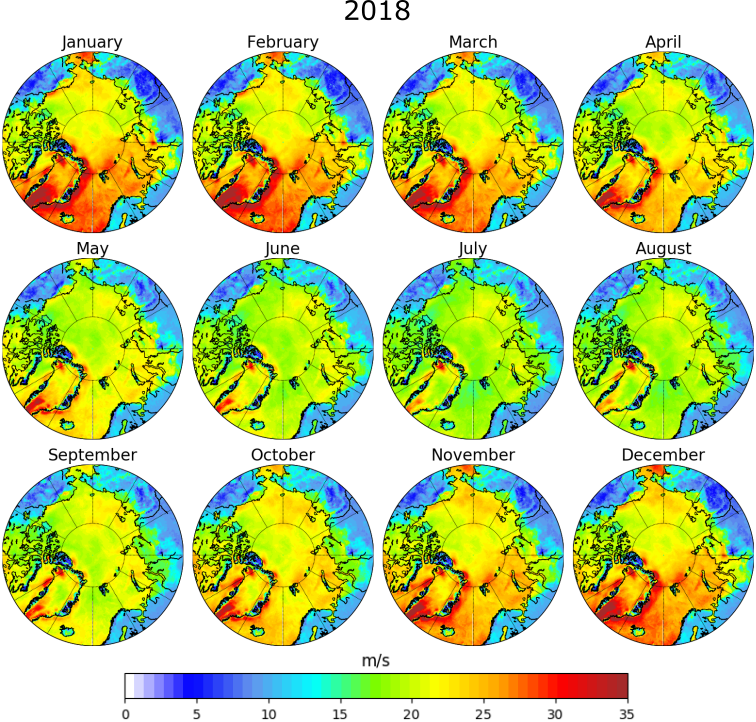}
  \caption{$U_{10}^{100}$ (m/s) for 2018 obtained with a POT analysis and a GPD distribution in the TS-EVA seasonal approach. Data obtained from the ERA5 dataset.}
  \label{u10-seasonal-2018}
\end{figure}

Figures \ref{u10-seasonal-1993} and \ref{u10-seasonal-2018} show the monthly values of $U_{10}^{100}$ for 1993 and 2018, respectively. Extreme wind distributes rather uniformly over the Arctic Ocean. During the autumn and winter season, $U_{10}^{100}$ ranges between 20 and 30 m/s, with peaks along Greenland (Denmark and Fram Straits) up to 50 m/s. In the spring and summer months, $U_{10}^{100}$ ranges between 10 and 30 m/s with again the highest winds reported in the western Greenland sea. Note that the seasonal approach returns a geographical distribution of extremes that is similar to the one obtained with the nonseasonal approach, but it captures more extreme season-related events. The seasonal component tends to shift the tail of the time-varying extreme value distribution into higher frequencies, resulting in higher estimated extremes for all seasons.

\begin{figure}
  \centering
  \includegraphics[scale=0.7]{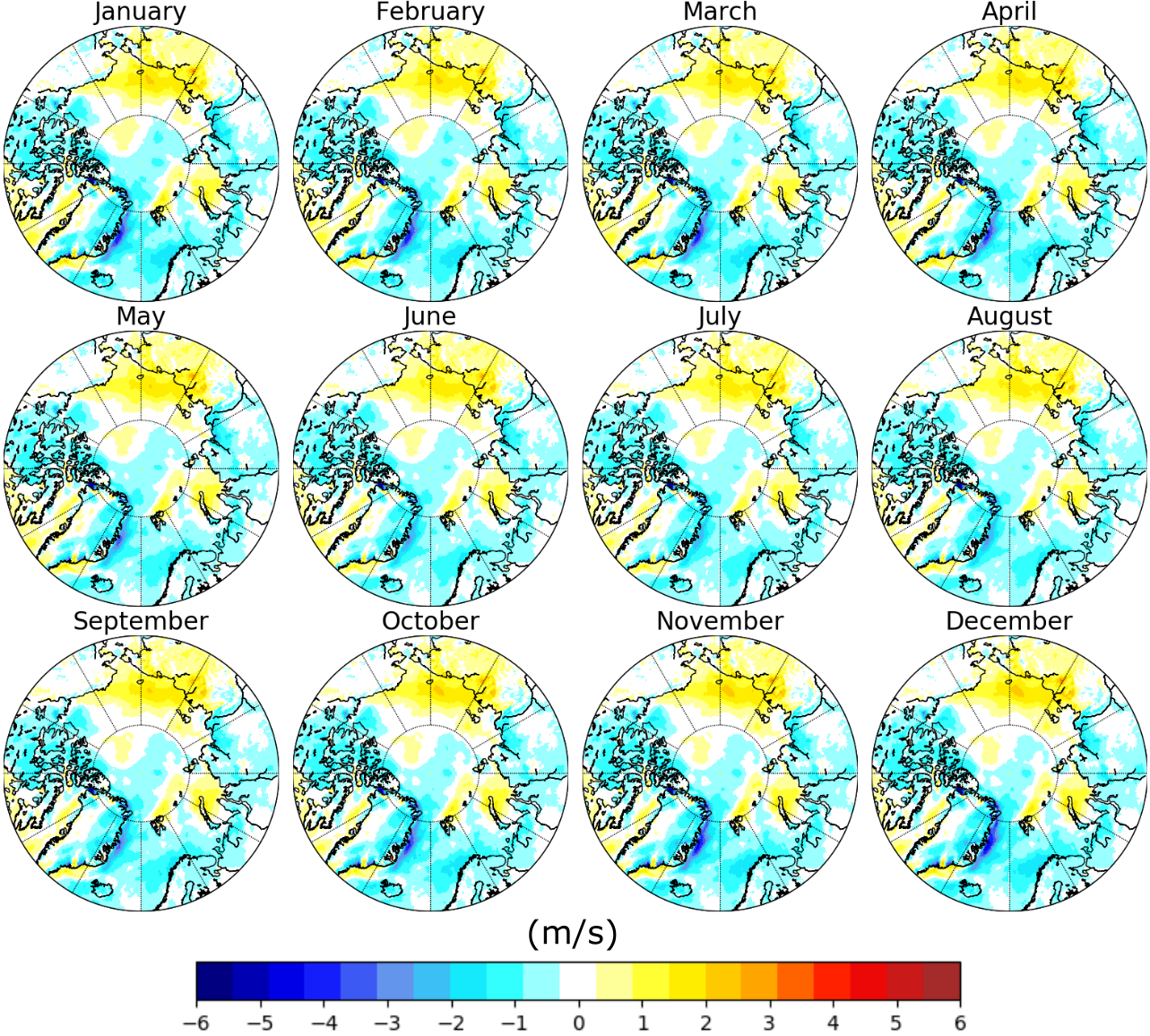}
  \caption{Monthly differences in $U_{10}^{100}$ between estimates for 2018 and 1993.}
  \label{seasonal-u10-1993-2018}
\end{figure}

Differences between $U_{10}^{100}$ for 1993 and 2018 are reported in Fig. \ref{seasonal-u10-1993-2018}. Generally, differences range between 1 and 3 m/s and are quite consistent across all seasons. The Pacific sector experiences an increase, while the Atlantic sector and the central Arctic are subjected to a reduction of $U_{10}^{100}$. The most significant changes are observed in the western Greenland sea during the winter season, where reductions up to -5 m/s were detected. It is interesting to note that the regional distribution of differences is similar in each month, denoting a homogeneous change of extreme winds across the Arctic Ocean throughout the year. Note also that differences obtained with the seasonal approach are consistent with those estimated with the nonseasonal method.

\subsection{Wave extremes}

\begin{figure}
  \centering
  \includegraphics[scale=0.7]{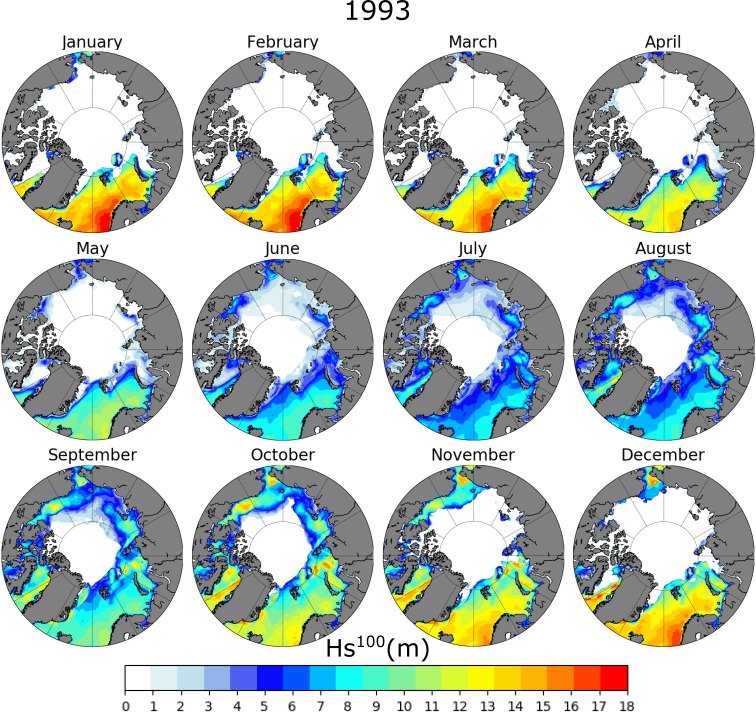}
  \caption{$H_s^{100}$ (m) for 1993 obtained with a POT analysis and a GPD distribution in the TS-EVA seasonal approach. Data obtained from the 28-year wave hindcast with ERA5 wind forcing.}
  \label{st6-seasonal-1993}
\end{figure}

\begin{figure}
  \centering
  \includegraphics[scale=0.7]{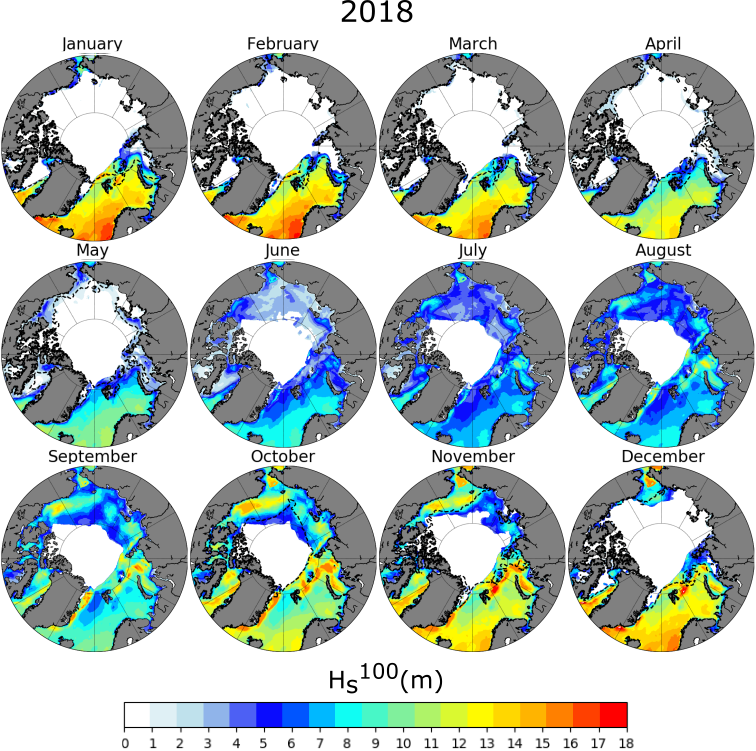}
  \caption{$H_s^{100}$ (m) for 2018 obtained with a POT analysis and a GPD distribution in the TS-EVA seasonal approach. Data obtained from the 28-year wave hindcast with ERA5 wind forcing. Dashed lines represent the minimum sea ice coverage in the period between 1991-1993 for each month.}
  \label{st6-seasonal-2018}
\end{figure}

\begin{figure}
  \centering
  \includegraphics[scale=0.7]{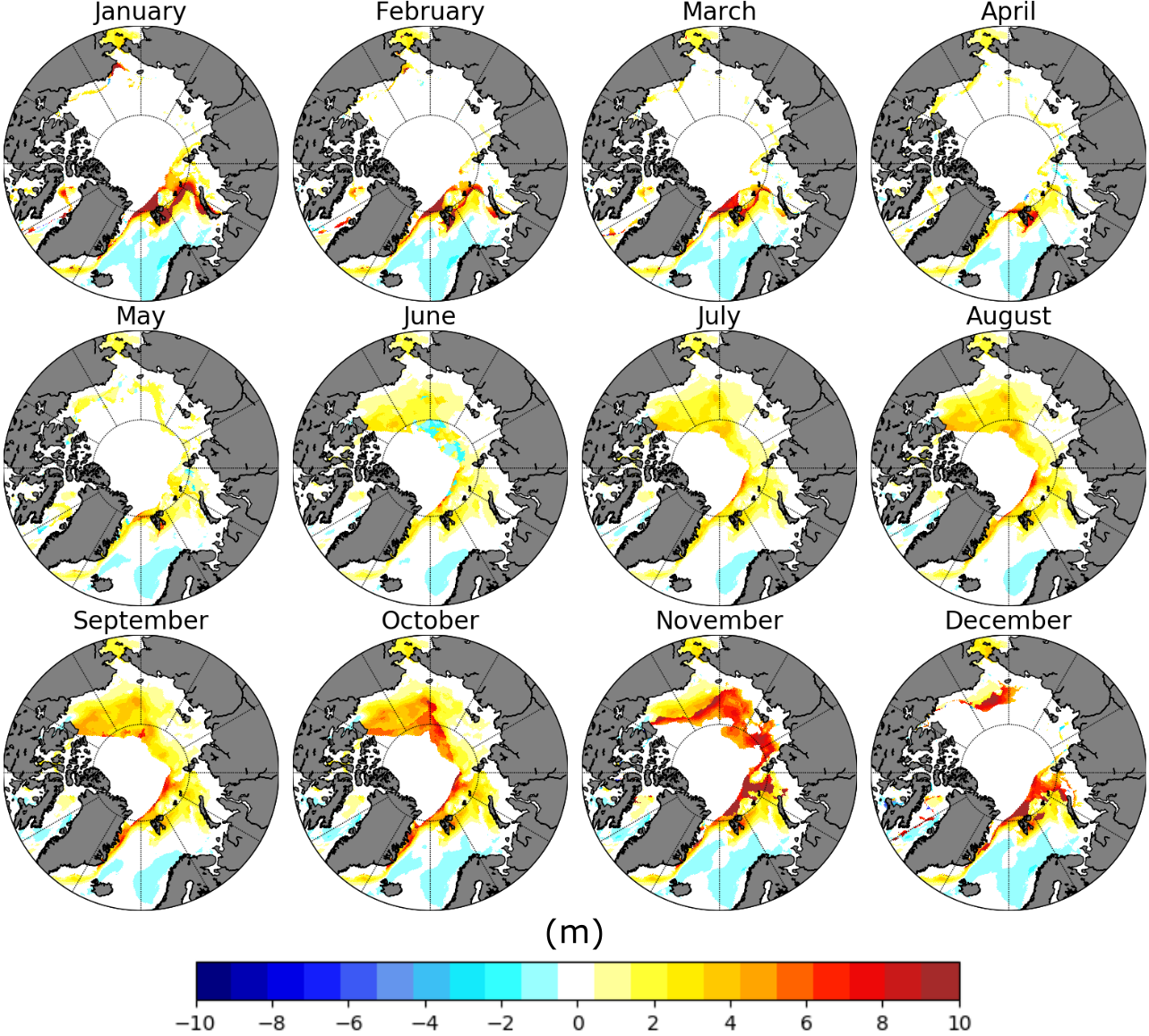}
  \caption{Monthly differences in $H_s^{100}$ between estimates for 2018 and 1993.}
  \label{seasonal-hs-1993-2018}
\end{figure}

\begin{figure}
  \centering
  \includegraphics[scale=0.18]{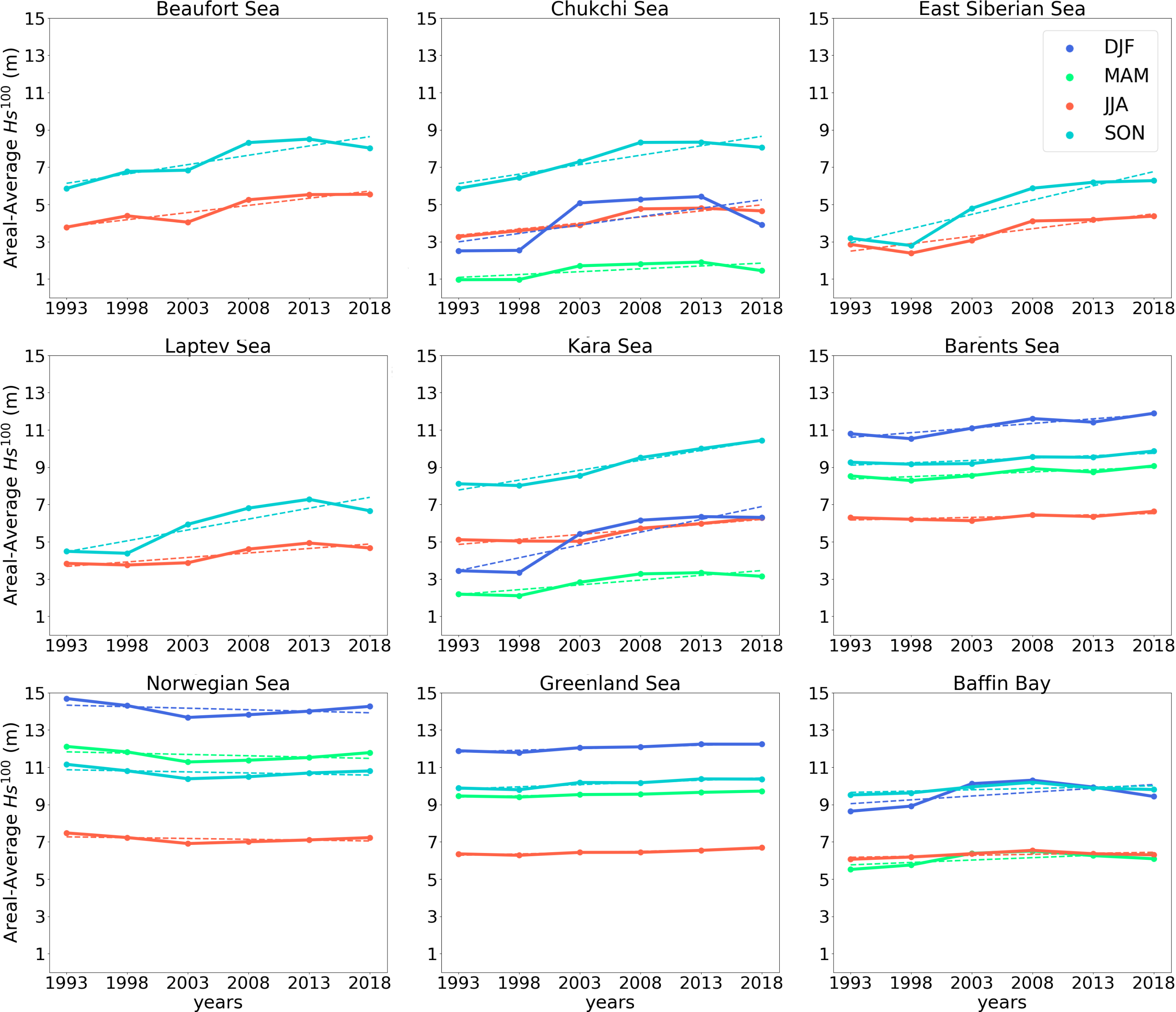}
  \caption{Areal-averages of $H_s^{100}$ in meters estimated by the seasonal TS-EVA approach for each sea in the Arctic Ocean for winter (blue), spring (light green), summer (red), and autumn (light blue).}
  \label{seasonal-area-hs}
\end{figure}

The seasonal variations of $H_s^{100}$ are presented in figures \ref{st6-seasonal-1993} and \ref{st6-seasonal-2018} for 1993 and 2018, respectively. The minimum sea ice coverage in 1991-1993 is shown as a dashed lines in Fig. \ref{st6-seasonal-2018}.
Extreme wave height, as expected, is subjected to a substantial seasonal variation. The highest values are found in the region encompassing the Greenland and Norwegian Seas, where energetic swells coming from the North Atlantic Ocean propagate into the region \citep[cf.][]{Liu2016,Stopa2016a}. The highest $H_s^{100}$ in this region reaches values up to 18 m in the winter months, concomitantly with strong winds (Figs. \ref{u10-seasonal-1993} and \ref{u10-seasonal-2018}), and reduces to about 5\;m in the summer. Over the past three decades, however, the general trend shows a consistent reduction in this region at a rate of 4 cm per year regardless of the season (see maps of differences in Fig. \ref{seasonal-hs-1993-2018} and trends of areal-averages in Fig. \ref{seasonal-area-hs}). These results are in agreement with the results obtained with the nonseasonal approach. Nevertheless, extreme waves penetrate further North in the emerging open waters of the Northern Greenland, Barents and Kara seas, especially during the autumn (September to November) and winter (December to February) seasons in recent years. Consequently, there is a dramatic increase of $H_s^{100}$ in these regions with values up to 13 m in 2018. This corresponds to an average increasing rate of approximately 12 cm per year, with peaks of about 35 cm per year nearby the sea ice margins. Based on future projection, this positive trend is expected to continue \citep{aarnes2017projected}.    

In regions subjected to the sea ice cycle, wave extremes in 1993 used to build up in late spring or early summer (June), and reach their maximum of up to 12 m in a confined area of the Beaufort sea in October. In more recent years (2018), waves already have a significant presence earlier in spring (May), primarily in the coastal waters of the Beaufort sea and the East Siberian sea (see figures \ref{seasonal-hs-1993-2018}). From June to November, there is a rapid intensification of the sea state and extremes span from a few metres in June to about 16 m in November, with an average growth rate of 12 cm per year, over a region encompassing the whole Beaufort, Chukchi and East Siberian seas. These secluded areas, which are the most prone to positive long-term variations of wind speed (Fig. \ref{seasonal-u10-1993-2018}) and sea ice retreat \citep{strong2013arctic}, are now experiencing sea state extremes comparable to those reported in the North Atlantic. It is also worth noting that significant changes are also apparent for the western part of the East Siberian sea and the nearby Laptev sea at the end of autumn (November). These regions, which used to be entirely covered by sea ice by November in the earliest decade, are now still completely open with $H_s^{100}$ recording changes up to 8 m (a rate of 32 cm per year since 1993).


\section{Conclusions} \label{conc}

A non-stationary extreme value analysis \citep[TS-EVA,][]{Mentaschi2016} was applied to assess long-term and seasonal variability of wind and wave extremes (100-year return period levels) in the Arctic Ocean. This non-conventional approach is dictated by the highly dynamic nature of the Arctic, which has been undergoing profound changes over the past decades \citep{Liu2016,Stopa2016a} and invalidating the basic hypothesis of stationarity that is fundamental for classical extreme value analysis. Estimation of extremes was based on a 28-year (1991-2018) database of 10-metre wind speed and significant wave height, with a temporal resolution of three hours. Wind speed was obtained from the recently released ERA5 reanalysis database and subsequently used to force the WAVEWATCH III spectral wave model. An Arctic Polar Stereographic Projection grid with a horizontal resolution spanning from 9 to 22 km was applied. The model was calibrated and validated against satellite altimeter observations, producing good agreement with a correlation coefficient $R = 0.97$, scatter index $SI = 16\%$ and root mean squared error $RMSE = 0.36$ m. 

The TS-EVA extreme value analysis consisted of transforming the original non-stationary time series of wind speed and wave height into a stationary counterpart and then applying standard peak-over-threshold methods to evaluate extreme values with a return period of 100 years over a running window of 5 years. Non-stationarity was then reinstated by back-transforming the resulting extreme value distribution. Two different approaches were applied to the data sets: a nonseasonal approach, which returns yearly estimates of extremes and enables evaluation of long-term variability; and a seasonal approach, which incorporates a seasonal variability enabling estimation of extremes for specific months.

The nonseasonal approach showed a weak long-term variability for the 100-year return period values of wind speed. An increase of approximately 3 m/s from 1993 to 2018 (a rate of $\approx$ 0.12 m/s per year since 1993) was reported in the Pacific sector, especially in the regions of the Chukchi and East Siberian seas and, more marginally, in the Beaufort sea and part of the Laptev sea. A decrease of roughly 3 m/s ($-0.12$ m/s per year), on the other hand, was found in most of the remaining regions of the Arctic, with peaks in the Eastern part of the Greenland sea ($\approx$ $-0.2$ m/s per year). Variability of wave extremes, on the other hand, is more dramatic and primarily driven by the substantially longer fetches following sea ice retreat. Large changes, in this respect, were found in the Pacific sector encompassing the area between the Beaufort and East Siberian seas, where wave height extremes have been increasing at a rate of approximately 12 cm per year, which results in an overall increase of $\approx$ 60\% from 1993 to 2018. It is interesting to note that wind extremes in the Beaufort sea only increased marginally, reinforcing the role of the sea ice decline in changing wave climate. The Atlantic sector, on the contrary, experienced a notable decrease of wave extremes at the rate of -4 cm per year; this is consistent with a reduction of wind extremes and with general climate trends observed in \citet{Liu2016}. For regions closer to the sea ice edge, where emerging open waters have been replacing pack ice, the 100-year return period levels of wave height exhibit the opposing trend, with a sharp increase of wave extremes at an extremely large local rate of 35 cm per year. It should be noted, however, that estimates of long-term trends closer to the sea ice edge are more uncertain due to lack of data in the earlier years, where sea ice covered the ocean more substantially. Nevertheless, it is worth reflecting on the consequences that a sharp upward trend of wave extremes can have on already weak sea ice. As extremes become more extreme, there is negative feedback accelerating sea ice dynamics \citep{vichi2019effects,alberello2020jgr}, break up \citep{passerotti2020omae} and melting processes \citep{dolatshah2018hydroelastic}, further contributing to sea ice retreat.   

The seasonal approach provides a more detailed picture of climate, providing a combined seasonal and long-term variability. Wind extremes distribute uniformly over the Arctic, with peaks in the autumn and winter periods spanning from 20 m/s in the Pacific sector to 30 m/s in the North Atlantic. Spring and Summer months still exhibit significant extremes up to 20 m/s, with a more homogeneous regional distribution. Over the entire 28-year period, trends are mild and stable through the seasons, consistent with those found with the nonseasonal approach. Variability of wave extremes is again more substantial than wind. In the Pacific sector, the decline of sea ice extent allows a rapid intensification of extremes in the spring (May and June); average growth rates span from 1 cm per year in spring to 12 cm per year in late summer and early autumn. In the Atlantic sector, in response to a notable drop of wind speed, a consistent decrease of wave extremes results all year-round. Nevertheless, the emerging waters of northern Greenland and Barents sea showed the opposite trend with an increase of wave height at a very large rate up to 32 cm per year closer to the sea ice margin.

{\bf Acknowledgments} 
This research was partially supported by the Victoria Latin America Doctoral Scholarship (VLADS) program. AT acknowledges support from the ACE Foundation--Ferring Pharmaceuticals and the Air-Sea-Lab Project initiative.

\section*{Appendix A: ERA5 wave data comparison} \label{A}

\begin{figure}
 \centerline{\includegraphics[scale=0.49]{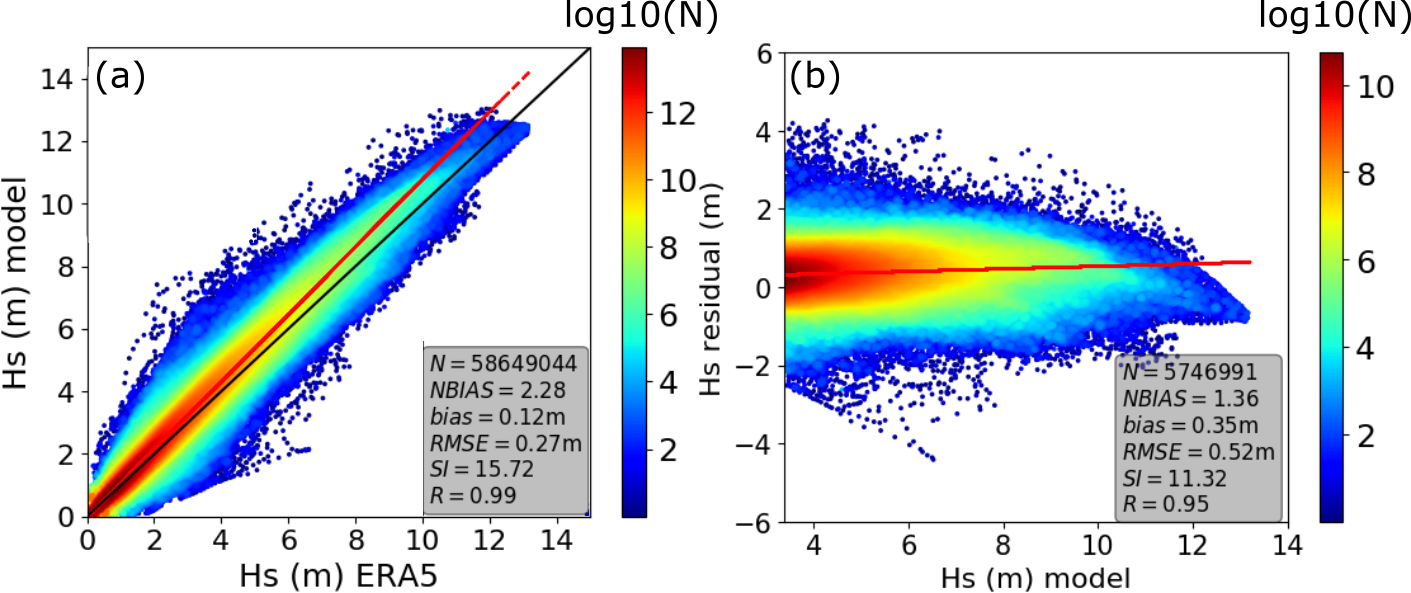}}
\caption{A1}{Significant wave height from ERA5 reanalysis versus WW3 model results for 2015. (a) all data and (b) $90^{th}$ percentile and above. The black line represents the 1:1 agreement and the red lines are the linear regression.} \label{era5mod}
\end{figure}

In addition to comparing the model results against altimeter data as described in section \ref{ww3}, an evaluation of the wave hindcast was also performed against the ERA5 wave data reanalysis. Although the ERA5 reanalysis has a coarser spatial resolution ($0.5^{\circ}$) than the hindcast performed in this study, it is a widely used international resource. The comparison between the WW3 wave hindcast and ERA5 data for 2015 (Fig. A1) shows excellent agreement with a correlation (R) of 0.99, RMSE of 0.27 m NBIAS of $2.3\%$ for all data and correlation (R) of 0.95, RMSE of 0.52 m NBIAS of $1.4\%$ for the upper percentiles.

 \bibliographystyle{ametsoc2014}

\end{document}